\documentclass[iop,apj,twocolumn,numberedappendix,twocolappendix]{openjournal}

\usepackage[utf8]{inputenc}
\usepackage{graphicx}	
\usepackage{amsmath}	
\usepackage{amssymb}	
\usepackage{orcidlink}

\usepackage[T1]{fontenc}

\usepackage{hyperref}
\hypersetup{
   colorlinks=true,
   linkcolor=blue,
   citecolor=blue,
   filecolor=magenta,      
   urlcolor=cyan,
   pdfpagemode=FullScreen,
   }

\makeatletter 
  \patchcmd{\NAT@citex}
    {\@citea\NAT@hyper@{%
      \NAT@nmfmt{\NAT@nm}%
      \hyper@natlinkbreak{\NAT@aysep\NAT@spacechar}{\@citeb\@extra@b@citeb}%
      \NAT@date}}
    {\@citea\NAT@nmfmt{\NAT@nm}%
    \NAT@aysep\NAT@spacechar\NAT@hyper@{\NAT@date}}{}{}

  \patchcmd{\NAT@citex}
    {\@citea\NAT@hyper@{%
      \NAT@nmfmt{\NAT@nm}%
      \hyper@natlinkbreak{\NAT@spacechar\NAT@@open\if*#1*\else#1\NAT@spacechar\fi}%
        {\@citeb\@extra@b@citeb}%
      \NAT@date}}
    {\@citea\NAT@nmfmt{\NAT@nm}%
    \NAT@spacechar\NAT@@open\if*#1*\else#1\NAT@spacechar\fi\NAT@hyper@{\NAT@date}}
    {}{}
\makeatother

\begin{document}
\title{Persephone's Torch: A 15th Magnitude Quadruply-Lensed Quasar From the Couch Discovered with SPHEREx and the LBT\vspace{-1.7cm}}
\author{Frederick~B.~Davies$^{1,\star}$$\orcidlink{0000-0003-0821-3644}$,     
Eduardo~Ba\~nados${^1}$$\orcidlink{0000-0002-2931-7824}$,
Sarah~E.~I.~Bosman$^{2,1}$$\orcidlink{0000-0001-8582-7012}$,
Arpita Ganguly${^2}$$\orcidlink{0009-0008-0444-4289}$,  
Silvia Belladitta${^{1,3}}$$\orcidlink{0000-0003-4747-4484}$, 
Jennifer Power${}^4$$\orcidlink{0009-0006-3761-3881}$ 
\& Jon Rees${}^4$$\orcidlink{0000-0002-5376-3883}$
\vspace{0.2cm}}
\thanks{$^{\star}$ Email: \href{mailto:davies@mpia.de}{davies@mpia.de}}

\affiliation{$^{1}$Max-Planck-Institut f\"ur Astronomie, K\"onigstuhl 17, D-69117, Heidelberg, Germany \\
$^{2}$Institute for Theoretical Physics, Heidelberg University, Philosophenweg 12, D–69120, Heidelberg, Germany \\
$^{3}$ INAF - Osservatorio di Astrofisica e Scienza dello Spazio di Bologna, Via Gobetti 93/3, I-40129 Bologna, Italy \\
$^{4}$Large Binocular Telescope Observatory, University of Arizona, 933 N Cherry Avenue, Tucson, AZ 85719, USA}

\begin{abstract}
\noindent Here we report the spectroscopic and geometric confirmation of an extremely bright ($i=14.77$) and compact (Einstein radius of $\sim0.45''$) quadruply-lensed quasar at $z=2.22$, 
J1330$-$0905, which we dub Persephone's Torch. The system had been previously selected as a candidate lensed quasar based on large-area survey data; here we confirm its quasar nature and redshift using public spectrophotometry from the SPHEREx mission, a.k.a.~``from the couch''. Adaptive optics imaging with LBT/LUCI resolves four images in a ``circular kite'' configuration. The system is the brightest gravitationally-lensed quasar system ever found. While an elliptical power-law mass distribution plus external shear accurately reproduces the locations of the images and lensing galaxy, and predicts a total magnification of $\sim56$, the brightnesses of the lensed images present highly anomalous flux ratios. Together with short time delays between images ($\leq 2$ days), this makes Persephone's Torch a promising candidate for future microlensing studies. Our discovery highlights the potential of SPHEREx full-sky infrared spectrophotometry to uncover extraordinarily bright objects that have otherwise been overlooked.
\end{abstract}


\maketitle

\section{Introduction}\label{sec:intro}

Strong gravitational lenses of quasars, and especially those producing four or more lensed images, serve a wide variety of purposes. Cosmological distances can be directly measured by employing time delays between their images (e.g.~\citealt{Treu16,Birrer24}) while they also provide multiple closely-separated sightlines through their foregrounds to precisely map the circumgalactic medium of lensing galaxies (e.g.~\citealt{Zahedy16}). Departures from the expected relative fluxes of the lens images (flux ratio anomalies) are used to detect and study small-scale structure in the background quasars (e.g.~\citealt{Kochanek04,Pooley07,DA11}), the lenses (e.g.~\citealt{MS98,Jimenez15}), and dark matter (e.g.~\citealt{DK02,Nierenberg24}). Comparison between the inferred gravitational and luminous masses of the lens galaxies also provide stringent tests for model of the initial mass function of stars (e.g.~\citealt{DAmato26}).
Despite their high usefulness, so-called ``quad'' lenses are very rare, with fewer than a hundred known over the entire sky \citep{Ducourant26}. The search for more quads is an active field of research (see e.g.~review by \citealt{Lemon24}). 

In \citet{Davies26QFTC}, we showed that The SpectroPhotometer for the History of the Universe, Epoch of Reionization, and Ices Explorer satellite (SPHEREx; \citealt{SPHEREx,Bock25} is an extremely useful tool for spectroscopic ``follow-up'' of targets 
without any additional telescope time -- that is, ``from the couch''. 
In that work, we used a simplistic color and astrometric selection for high-redshift quasars (similar to e.g.~\citealt{Yang17,Belladitta25}),
but much more sophisticated methods, aided by machine learning, have been successfully demonstrated in the literature (e.g.~\citealt{Schindler17,Nanni22,Guarneri22,Ye24,Byrne24,Calderone24,MartinezRamirez26}). 
Due to the difficulty in obtaining comprehensive spectroscopic follow-up,
many authors have published their candidate lists publicly (e.g. \citealt{Fu24,Ye24,Calderone24}).

\begin{figure*}
    \centering
    \includegraphics[width=0.375\linewidth]{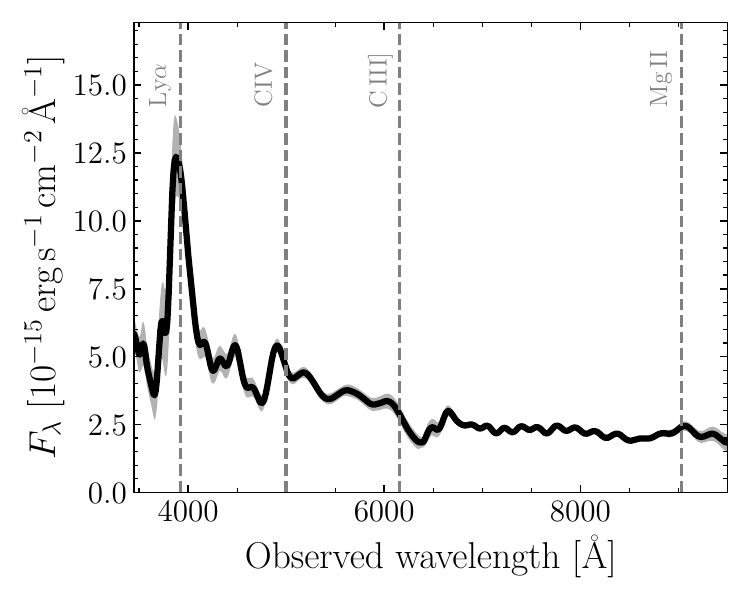}
    \includegraphics[width=0.6\linewidth]{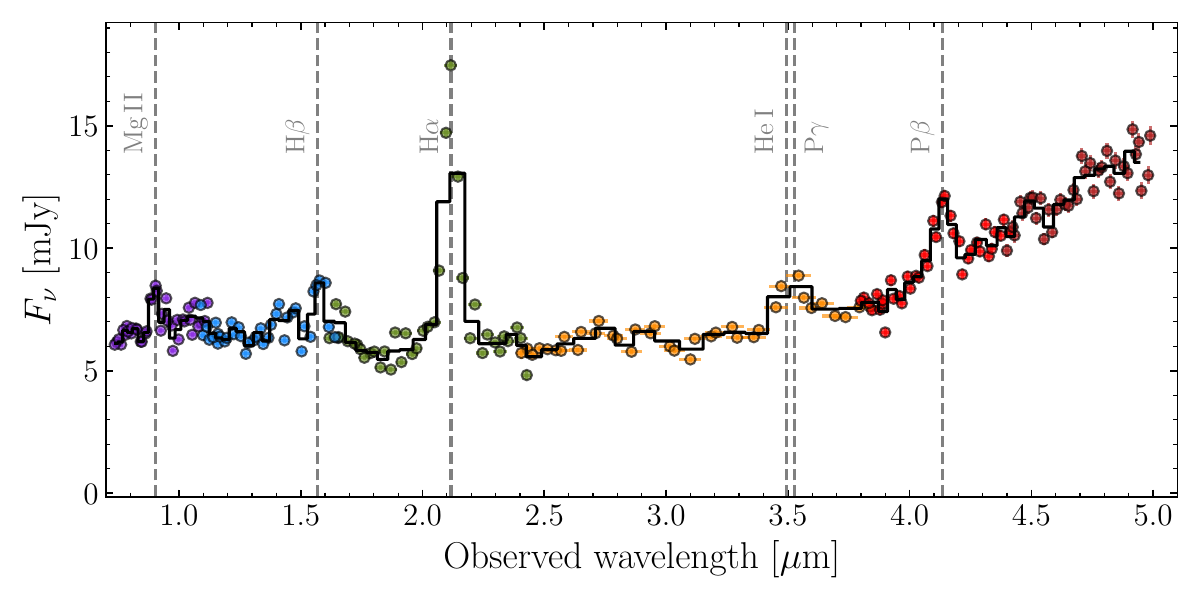}
    \vskip-0.5em
    \caption{\textit{Gaia} XP spectrum (left) and SPHEREx spectrophotometry (right) of J1330$-$0905. Common broad emission line features redshifted to $z=2.22$ are labeled with vertical dashed lines. }
    \label{fig:spec}
\end{figure*}

Here we show how SPHEREx can be used to test the bright end of public quasar candidate catalogs in a particularly striking fashion, resulting in the discovery of an extraordinarily bright quadruply-lensed quasar which we dub Persephone's Torch. Magnitudes are quoted in the AB magnitude system \citep{Oke83}, uncertainties are given at the $1\sigma$ level, angles are computed in the image plane a.k.a.~radians north of west, and the calculations of time delays assume a \textit{Planck} cosmology \citep{Planck18}.

\section{Confirmation of a peculiarly bright quasar candidate}\label{sec:spec}

Following the search for high-redshift quasars in \citet{Davies26QFTC}, we applied the SPHEREx Spectrophotometry Tool\footnote{\url{https://irsa.ipac.caltech.edu/applications/spherex/tool-spectrophotometry}}
to various public catalogs of high-redshift quasar candidates in the literature.
Further quasar discoveries (and refutations) will be described in future work, but here we discuss the brightest quasar confirmation that we have obtained thus far.

\citet{Calderone24} explored machine learning techniques to improve the selection
of bright quasars from public survey data (\textit{Gaia}, \citealt{Gaia}; WISE, \citealt{WISE,NeoWISE}; the Panoramic Survey Telescope and Rapid Response System (Pan-STARRS) survey, \citealt{Chambers16}) as part of the larger QUasars as BRIght beacons for Cosmology in the Southern hemisphere (QUBRICS) program \citep{Boutsia20,Guarneri22,Cristiani23,Porru26}.
By far the brightest candidate in their table B1 without spectroscopic confirmation is located at 13:30:05.26 $-$09:05:03.82, henceforth J1330$-$0905, with Pan-STARRS Kron magnitudes of $g=14.977\pm0.040$, $r=14.876\pm0.018$, $i=14.774\pm0.007$, $z=14.518\pm0.011$, and $y=14.537\pm0.011$.

We employed the SPHEREx Spectrophotometry Tool \citep{Akeson25} to extract the spectrum of J1330$-$0905 from the SPHEREx Quick Release data, using a multi-object catalog including the nearby ($\sim10$\,arcsec) 12th magnitude star \textit{Gaia} DR3 3629934525528245376 to deblend the two objects. The right panel of Figure \ref{fig:spec} shows the SPHEREx spectrum, identifying the object as an extremely bright quasar at $z = 2.2245 \pm 0.0005$ via several broad emission line features including H$\alpha$ and Paschen-$\beta$. The redshift is determined via a Gaussian fit to the H$\alpha$ line as in \citet{Davies26QFTC}. At the time we queried the data presented in Figure~\ref{fig:spec}, J1330$-$0905 had been observed by the first and second all-sky SPHEREx surveys approximately 6 months apart. In the second ``epoch'' (actually spread out over several days across the different spectral channels), the object was $\sim$\!$15$\% brighter than during the first ``epoch''; in Figure \ref{fig:spec} we have adjusted the measurements from the first epoch upward by this factor for clarity.

With a $g$-band magnitude of $\simeq15.0$, the implied absolute UV magnitude of $M_{1450}\approx-30$ would make J1330$-$0905 comparable to the most luminous quasar known at any redshift \citep{Wolf24} -- however, this far past the exponential turnover of the quasar luminosity function (e.g.~\citealt{Kulkarni19}), 
one must also account for the possibility of magnification by gravitational lensing \citep{Turner80}. 

\subsection{The lensing hypothesis}

In fact, \citet{Calderone24} was not the first to propose J1330$-$0905 as a candidate quasar. A previous work by \citet{Makarov23} included J1330$-$0905 on a list of candidate \emph{lensed} quasars\footnote{The trail actually goes back even further: the parent candidate quasar sample investigated by \citet{Makarov23}, including J1330$-$0905, was originally published by \citet{Secrest15}.}, due to its high value of the ``\texttt{phot\_bp\_rp\_excess\_factor}'' parameter in the \textit{Gaia} DR3 catalog. They also estimated a photometric redshift of $z_{\rm phot}\simeq2.08$, comparable to our spectroscopic redshift determination.

Other features of apparently isolated \textit{Gaia} sources have been used to identify candidate close pairs of objects in the literature, e.g.~the excess astrometric noise \citep{Hwang20} and the fraction of \textit{Gaia} epochs with multiple peaks detected \citep{Mannucci22}. In both cases, J1330$-$0905 is at the highest end of the distribution, with \texttt{astrometric\_excess\_noise} $=16.2$\,mas and \texttt{ipd\_frac\_multi\_peak} $=0.50$. Clearly, there is strong evidence for small-scale multiplicity, likely due to lensing.

\section{High-resolution imaging followup with LBT/LUCI}\label{sec:img}

\begin{figure*}
    \centering
    \includegraphics[width=0.99\linewidth]{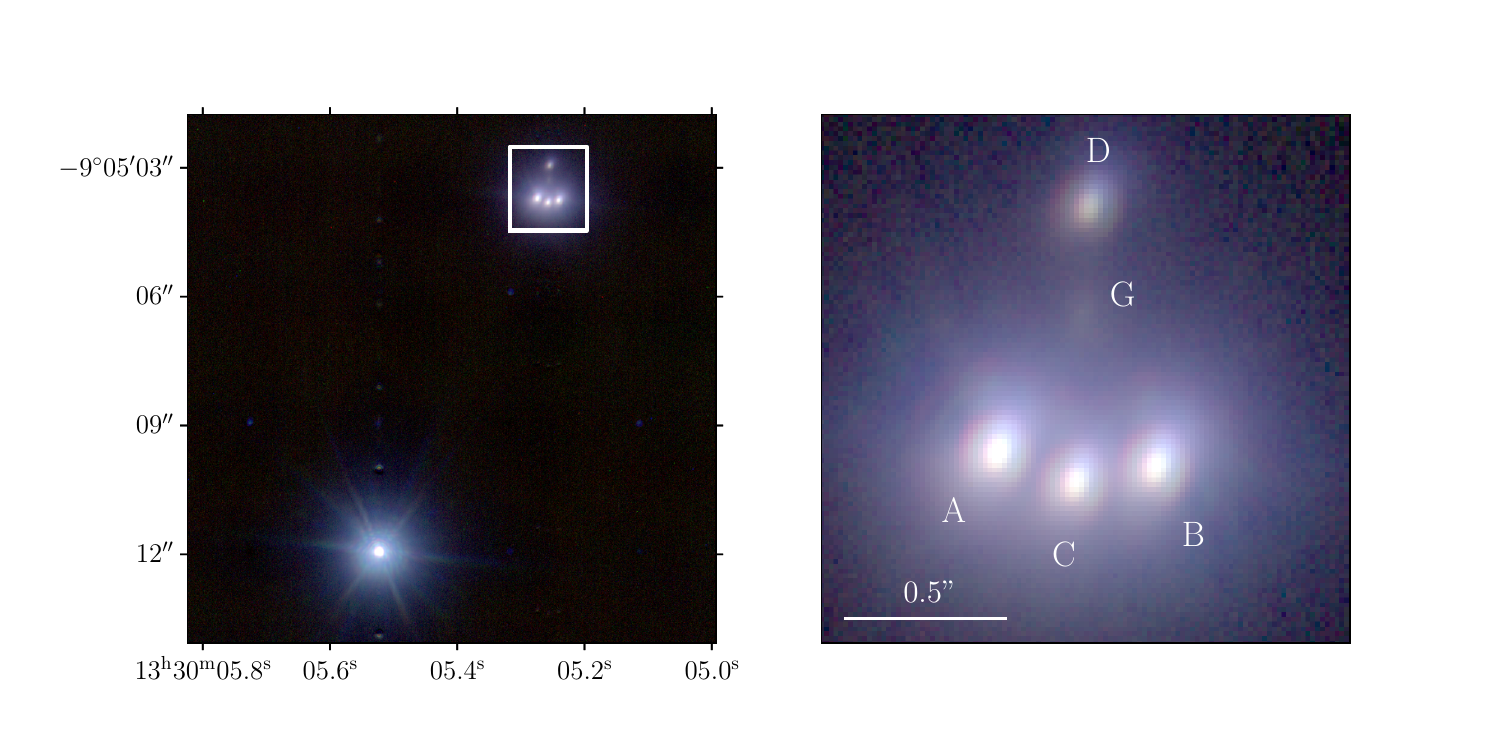}
    \vskip-2em
    \caption{Color-composite $JHK$ LBT/LUCI1 image of the J1330$-$0905 field (left) and zoomed-in cutout (right) with a logarithmic stretch to highlight faint features. We label the individual images ABCD from brightest (lower left) to faintest (top) along with the presumptive lens galaxy G (middle).}
    \label{fig:lbt}
\end{figure*}

To definitively resolve the question of lensing, on 26 March 2026 we obtained near-infrared imaging of J1330$-$0905 with the
LBT Utility Camera in the Infrared \citep[LUCI,][]{LUCI} on the Large Binocular Telescope (LBT) in the $J/H/K$ bands ($315/144/96$\,s total exposure time, respectively) using the SOUL adaptive optics (AO) system \citep{SOUL} during seeing conditions of $\sim0.7''$. We used the N30 camera designed for AO observations, with a pixel scale of $0.015''$ and $30''\times30''$ field of view. The nearby bright star, which originally complicated the SPHEREx spectrophotometry, was used for wavefront corrections. The exposures were taken in a star-shaped 9-point dither pattern with a characteristic scale of $3''$, median-subtracted, then combined by shifting to the offset corresponding to the maximum cross-correlation with a single reference exposure.

Figure \ref{fig:lbt} shows a colorized combined image from LUCI1, revealing clear hallmarks of a quadruple lens system -- three similarly bright, closely-spaced images in the south, one fainter image in the north, and plausible faint fuzz in-between, presumably from the lens galaxy. The data from LUCI2 displays a less well-behaved AO point-spread-function, and we do not analyze it in this work. Due to its location in the constellation of Virgo, and its extraordinary brightness, we dub this system Persephone's Torch\footnote{Drawing some inspiration from another extremely bright quadruple lensed quasar, Andromeda's Parachute \citep{Berghea17,Rubin18}.}. The combination of the four images, $i = 14.77$ in Pan-STARRS, is the brightest total magnitude of \textit{any} gravitationally-lensed quasar system ever found (comparable to Andromeda's Parachute with combined $i$-band magnitude $\sim$\!14.9, cf. \citealt{Schmidt23}). 

\subsection{Lens modeling}

\begin{figure}[t]
    \centering
    \includegraphics[width=0.95\linewidth]{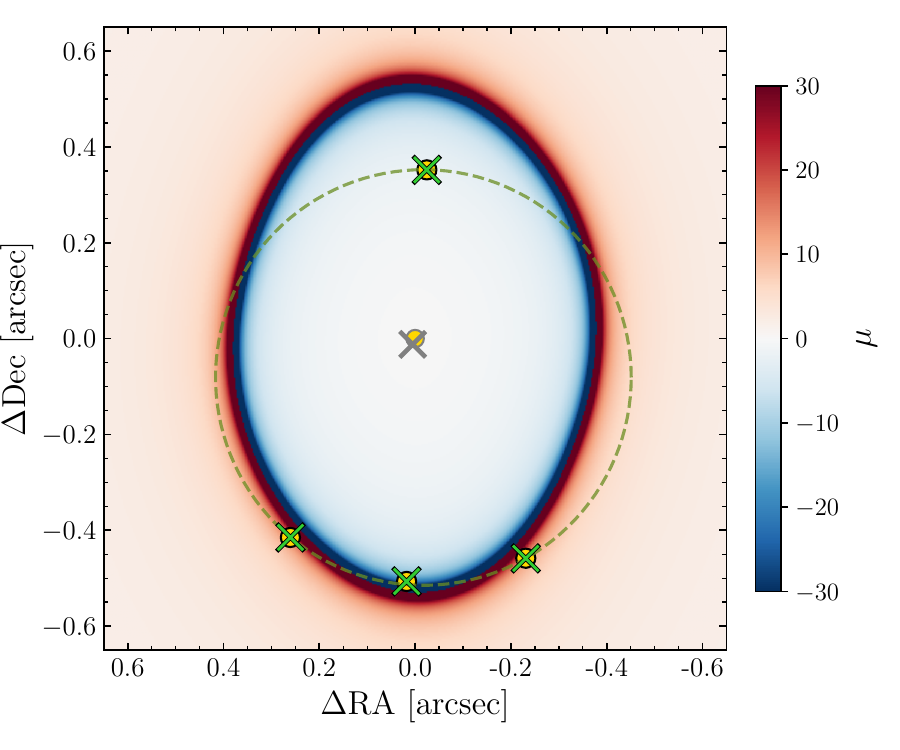}
    \vskip-0.5em
    \caption{Optimal lens model for the system. The circles display the measured centroids of the four quasar images while green crosses show their corresponding predicted locations; the grey cross shows the position of the lensing galaxy. The color bar indicates the magnification $\mu$. The dashed green circle demonstrates that all four images lie almost exactly on a common circle.}
     \vskip-1em
    \label{fig:lens}
\end{figure}

We fit a lensing model for the system using the \texttt{lenstronomy} code \citep{lenstronomy}. The lensing galaxy is modeled with an elliptical power-law mass distribution plus an external weak-lensing shear, as in e.g.~\citet{Shajib19}. We optimize the resulting ten parameters to reproduce the observed locations of the four lensed images with respect to the lensing galaxy, while allowing for an uncertainty in the centroid of the latter. The best-fit lens model, shown in Figure~\ref{fig:lens}, uses a lensing galaxy potential with ellipticity $e=0.626$, an Einstein radius of $0.446$\,arcsec, orientation angle $1.657$\,rad and density profile index $\gamma=2.011$. The external shear at the location of the system is best fit with an orientation of $1.907$\,rad and intensity $\gamma_\text{shear} = 0.075$. Finally, the optimal model prefers a position offset $\Delta({\rm RA},{\rm Dec})=(-0.0097,-0.0729)$ arcsec of the background source behind the lens, and an offset $\Delta({\rm RA},{\rm Dec})=(0.0052,0.012)$ arcsec of the center of the lensing potential compared to the measured centroid of the galaxy. This offset is unlikely to be physical; the lensing galaxy is faint and contaminated by the glare of the seeing disk from the three brightest lensed images, complicating its centroid determination. 

A search through the Strong Lensing Database (SLED\footnote{\url{https://sled.amnh.org/}}) suggests that the system's Einstein radius is among the ten smallest found in quadruply-lensed quasars to date. While the lensing galaxy redshift $z_\text{lens}$ is not known, the best-fit model predicts short time delays between the images whether we assume $z_\text{lens}=0.5$ ($\Delta t_\text{max}=16.5$ hours) or $z_\text{lens}=1.0$ ($\Delta t_\text{max}=2.0$ days).  
The resulting predicted magnifications $\mu$, relative image brightnesses, and time delays are given in Table~1.

The lensing system has a typical ``kite'' configuration, or more accurately, a long-axis cusp configuration (e.g.~\citealt{Hou26}) or a (perfectly) circular kite configuration, as we should with a green dashed circle in Figure~\ref{fig:lens} \citep{Schechter22}. In this set-up, the brightest image is always expected to be the one directly across from the faintest image, i.e.~image C (e.g.~\citealt{Keaton03}). Instead, the brightest image is A, with $F_A/F_C \sim 1.4$. Since this is not possible to match with straightforward lensing potentials, we do not attempt to. A departure from the expected brightness ordering of the images likely indicates small-scale structure in either the background quasar or the lensing galaxy, or alternatively, microlensing.

\section{Discussion \& Conclusion}\label{sec:disc}

\begin{table}[t!]
    \begin{tabular}{l | c c c c}
 & A & B & C & D \\  
 \hline \hline
$F_X/F_A$ & 1.00 & 0.74 & 0.70 & 0.08 \\
$\Delta{\rm RA}_{X-A}$ ('') & 0 & $-0.4911$ & $-0.2423$ & $-0.2845$\\
$\Delta{\rm Dec}_{X-A}$ ('') & 0 & $-0.0437$ & $-0.0918$ & $0.7672$\\
\hline 
$\mu$ & $16.42$ & $13.61$ & $-24.39$ & $-1.99$ \\
$|\mu_X/\mu_A|$ & 1.00 & 0.84 & 1.49 & 0.12 \\ 
$\Delta{\rm RA}_{X-A}$ ('') & 0 &  $-0.4913$ & $-0.2425$ & $-0.2848$\\
$\Delta{\rm Dec}_{X-A}$ ('') & 0 & $-0.0433$ & $-0.0918$ & $0.7673$\\
$\Delta t_{AX}$, $z_\text{lens} = 0.5$ (d) & --- & $0.16$ & $-0.53$ & $-0.27$ \\
$\Delta t_{AX}$, $z_\text{lens} = 1.0$ (d) & --- & $0.45$ & $-1.52$ & $-0.78$ \\
\hline \hline
    \end{tabular}
    \vskip 1em
    Table 1. Properties of the four lensed images. Quantities above the horizontal line are directly measured, while the ones below are inferred from the best-fit lensing model. The redshift of the lensing galaxy is not known; we list the time delays corresponding to $z_\text{lens}=0.5, 1.0$ for reference.

\end{table}

Here we have reported the spectroscopic and high-resolution imaging confirmation of the extremely bright quadruply lensed $z=2.22$ quasar J1330$-$0905, a.k.a.~Persephone's Torch. We applied the SPHEREx Spectrophotometry Tool to the brightest quasar candidate in \citet{Calderone24} to confirm its quasar nature and redshift, then obtained LBT/LUCI AO imaging observations to characterize its small-scale multiplicity indicated by \emph{Gaia} \citep{Makarov23}. Persephone's Torch is virtually tied with Andromeda's Parachute as the brightest gravitationally-lensed system known to date at any redshift.

Given its extreme brightness and relatively inconspicuous location on the sky (i.e. not close to the Galactic plane or any particularly bright star or external galaxy), one may ask why J1330$-$0905 was not confirmed previously. The coordinates were covered by the objective prism imaging of the Hamburg/ESO Survey (HES), which in principle should have confirmed its redshift during the 1990s \citep{Wisotzki00}. However, a query to the HES spectral archive \citep{HES_spectra} returns only the nearby bright star, flagged as  ``extended'' and with a hint of Ly$\alpha$ 1215\,\AA emission at $z=2.22$. This suggests that the two objects were blended, which may have obscured the quasar nature of the source.

While we used SPHEREx for confirmation, the left panel of Figure \ref{fig:spec} shows that its \textit{Gaia} XP spectrum from \textit{Gaia} DR3 would already support identification as a $z\sim2$ quasar via the strong Ly$\alpha$ emission line, and indeed it is classified by the \textit{Gaia} DR3 pipeline as a quasar with 99.7\% probability. However, we note that, likely due to the object's complicated structure on small scales, \textit{Gaia} DR3 does not provide a proper motion or parallax, i.e.~they are equal to ``\texttt{Null}'' in the catalog. The vast majority of quasar searches employ astrometric cuts to remove contamination by Galactic stars, which is especially likely at the bright apparent magnitude of J1330$-$0905, but this would leave out sources like J1330$-$0905. Interestingly, the \textit{Gaia} DR2 (but not DR3) catalog contains an additional faint source at the location of image D.

\begin{figure}[t]
    \centering
    \includegraphics[width=0.95\linewidth]{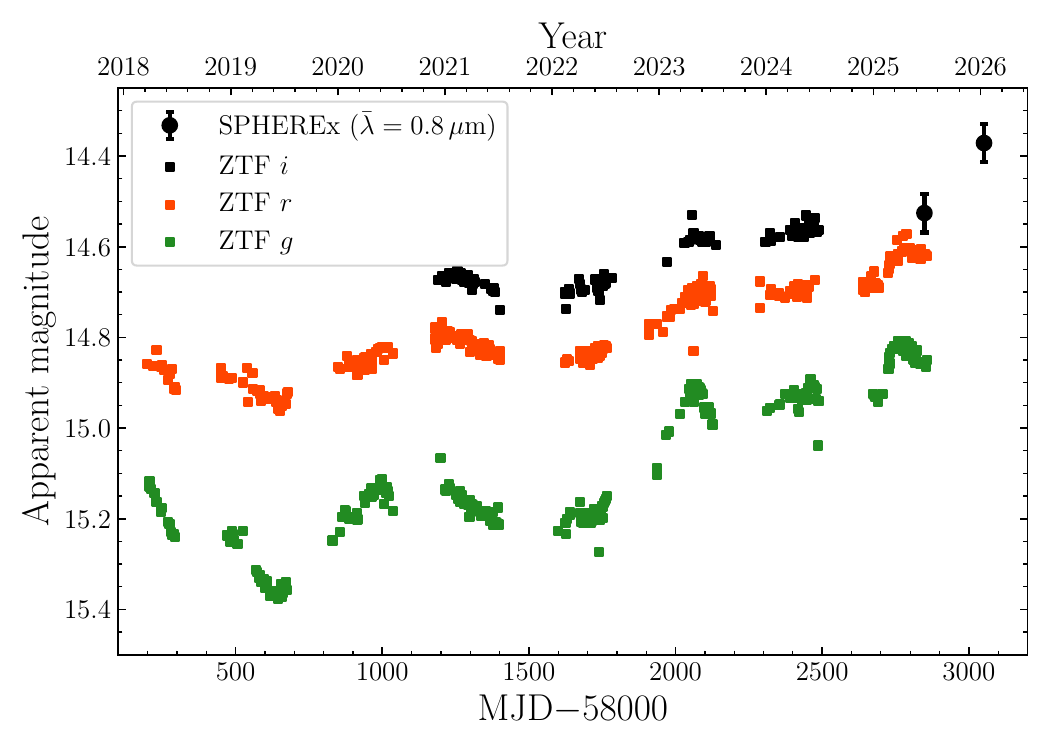}
    \vskip-0.5em
    \caption{Light curve of J1330$-$0905 from ZTF in $i$-band (black), $r$-band (red), and $g$-band (green). More recent photometry from SPHEREx is binned to match the $i$-band (points with uncertainties). The measurements include all four lensed images.}
    \vskip-1em
    \label{fig:ztf}
\end{figure}

Another piece of evidence which could have been used to support the quasar nature of J1330$-$0905 is its variability. In Figure \ref{fig:ztf} we show its 2018--2025 light curve from the most recent public data release (DR24) of the Zwicky Transient Facility \citep[ZTF;][]{ZTF1, ZTF2}, which shows the random walk pattern and chromaticity characteristic of quasars (e.g.~\citealt{Giveon99,Kelly09,Schmidt12}). Persephone's Torch was somewhat brighter in 2025 compared to its average during the Pan-STARRS survey (2009--2013). We also show the mean brightness from the two survey epochs of SPHEREx, averaging over a wavelength range comparable to the ZTF $i$-band ($0.736 < \lambda_{\rm obs} < 0.87$\,$\mu$m). As described in Section~\ref{sec:spec}, the second survey epoch is somewhat brighter than the first, and comparison to its historical light curve suggests that Persephone's Torch is currently at its brightest in recent history.

Persephone's Torch is also detected across the electromagnetic spectrum, providing additional characterization of the system. It is detected in the radio bands by the Rapid ASKAP Continuum Survey \citep[RACS,][]{mcconnell2020,hale2021} at 0.888, 1.37, and 1.67\,GHz, as well as by the VLA Sky Survey \citep[VLASS,][]{lacy2020} at 3\,GHz. The source is relatively faint, exhibiting a peak flux density of $\sim 1.6$\,mJy at 1.37, 1.67, and 3\,GHz, which suggests a flat spectrum at higher frequencies. 
In contrast, a higher peak flux density of $\sim 3$\,mJy at 0.888\,GHz indicates the presence of a steep-spectrum radio component (e.g., lobe emission) that becomes more prominent toward lower frequencies. 
Furthermore, the quasar has a clear detection at X-ray frequencies in the eROSITA-DE Data Release 1 (DR1, \citealt{Merloni24}). The source is identified as 1eRASS J133005.3$-$090504, and the separation between optical and eRASS coordinate is 1.9'' (i.e.~well within the 16'' PSF of eROSITA). The catalog reports a 0.2$-$2.3 keV X-ray flux of 1.17$\pm$0.31 $\times$ 10$^{-13}$ erg s$^{-1}$ cm$^{-2}$ (see \citealt{Merloni24} for more details).  

Our discovery shows the potential for public spectrophotometric surveys like SPHEREx to identify extraordinary objects that have been missed by previous surveys. While the close alignment of the predicted time delays between the four images complicates its use for sensitive time-delay cosmography, the significant flux ratio anomalies between the three bright images suggest Persephone's Torch may still be a powerful probe of small-scale structure, either from microlensing by stars within the lens galaxy (e.g.~\citealt{Wambsganss90,Vernardos24}) or from dark matter substructure along the line of sight (e.g.~\citealt{DK02,Vegetti24}).

\section*{Acknowledgements}
\small

The authors thank Cameron Lemon for providing access to SLED.

SEIB and AG are supported by the Deutsche Forschungsgemeinschaft (DFG) under Emmy Noether grant number BO 5771/1-1.

This work made use of \texttt{astropy} \citep{astropy}, \texttt{numpy} \citep{numpy}, \texttt{matplotlib} \citep{matplotlib}, \texttt{ipython} \citep{ipython}, \texttt{scipy} \citep{scipy} and \texttt{emcee} \citep{emcee}.

This research has made use of ``Aladin sky atlas'' developed at CDS, Strasbourg Observatory, France \citep{aladin}.

This work has made use of the Python package GaiaXPy, developed and maintained by members of the \textit{Gaia} Data Processing and Analysis Consortium (DPAC), and in particular, Coordination Unit 5 (CU5), and the Data Processing Centre located at the Institute of Astronomy, Cambridge, UK (DPCI).

This research has made use of the NASA/IPAC Infrared Science Archive, which is funded by the National Aeronautics and Space Administration and operated by the California Institute of Technology.

This paper includes data from the LBT (Program ID: MPIA-2026A-001).
The LBT is an international collaboration among institutions in the United States and Europe. At the time data were acquired for this research, LBT Corporation Members were the University of Arizona on behalf of the Arizona Board of Regents; Istituto Nazionale di Astrofisica, Italy; and The Ohio State University, representing OSU, University of Notre Dame, University of Minnesota, and University of Virginia. This research used the facilities of the Italian Center for Astronomical Archives (IA2) operated by INAF at the Astronomical Observatory of Trieste. Observations have benefited from the use of ALTA Center (\url{alta.arcetri.inaf.it}) forecasts performed with the Astro-Meso-Nh model. Initialization data of the ALTA automatic forecast system come from the General Circulation Model (HRES) of the European Centre for Medium Range Weather Forecasts.

    This work has made use of data from the European Space Agency (ESA) mission
{\it Gaia} (\url{https://www.cosmos.esa.int/gaia}), processed by the {\it Gaia}
Data Processing and Analysis Consortium (DPAC,
\url{https://www.cosmos.esa.int/web/gaia/dpac/consortium}). Funding for the DPAC
has been provided by national institutions, in particular the institutions
participating in the {\it Gaia} Multilateral Agreement.

The Zwicky Transient Facility is by the National Science Foundation under Grants No. AST-1440341 and AST-2034437 and a collaboration including current partners Caltech, IPAC, the Oskar Klein Center at Stockholm University, the University of Maryland, University of California, Berkeley, the University of Wisconsin at Milwaukee, University of Warwick, Ruhr University, Cornell University, Northwestern University and Drexel University. Operations are conducted by COO, IPAC, and UW.

SLED is supported by the American Museum of Natural History that kindly provides web hosting, resources, and infrastructure support, with particular thanks to Sajesh Singh, Lawrence Levinson, and Paul Delong from the AMNH IT team. SLED utilizes services provided by the AMNH ScienceDMZ. The AMNH Science DMZ and associated research computing services are supported by the National Science Foundation Campus Cyberinfrastructure (NSF CC$\star$) Awards 1827153, 1925590, and 2232857. 

\bibliographystyle{aasjournal}
 \newcommand{\noop}[1]{}

\end{document}